\pdfoutput=1
\documentclass[preprint,12pt]{elsarticle}
\usepackage{amsmath}
\usepackage{booktabs}
\usepackage{array}
\usepackage{graphicx}
\usepackage[hidelinks]{hyperref}
\journal{Journal of Systems and Software}

\begin{document}

\begin{frontmatter}

\title{MCP Error Messages Written for Developers Hurt the Most Capable Agents Most}

\author[gt]{Xiaonan Xu\corref{cor1}}
\ead{xiaonanxu5@gmail.com}
\author[cu]{Wenjing Wu}
\ead{wuwenjing256@gmail.com}
\affiliation[gt]{organization={College of Computing, Georgia Institute of Technology}, city={Atlanta}, postcode={GA 30332}, country={USA}}
\affiliation[cu]{organization={Department of Computer Science, University of Colorado Boulder}, city={Boulder}, postcode={CO 80309}, country={USA}}
\cortext[cor1]{Corresponding author.}

\begin{abstract}
Many Model Context Protocol (MCP) servers wrap web APIs built for human developers, and their error messages tell the reader to run a command, edit a configuration, open a web page or wait. Many agents that read them can only call the server's tools. In 150 widely used MCP servers, 949 of 3,001 error messages tell the caller what to do next, and half of these steps depend on something the server cannot see about the caller. On credential errors, 62 of 67 steps ask for a terminal command, a configuration change or a web page; on rate limits, 20 of 30 say to wait and retry without naming the call to repeat. We tested five OpenAI models that act only through the tools of Berkeley Function Calling Leaderboard tasks, and the agents did what the step said. On expired credentials, a terminal command in the step left 45\% of tasks recovered, and the loss it caused grew from 18 points for GPT-5.5 to 69 for GPT-6 Astra. On a rate limit, GitHub's ``Wait before retrying.'' left 6\%. We tested two remedies. For MCP developers, naming a server tool in the step raised recovery on expired credentials to 84\%, with the login tool in place of the command, and on a rate limit to 88\%, with the call to repeat in place of the bare wait. For agent developers, deleting the step with a one-sentence prompt before the model reads it raised recovery on expired credentials to 82\%.
\end{abstract}


\end{frontmatter}

\section{Introduction}

Web APIs were built for human developers, and their error messages are written for a reader who can act outside the failed call: open a terminal, edit a configuration file, sign in on a web page or wait a minute before trying again. ``Please run: \texttt{reddit-mcp-buddy -{}-auth}'' assumes a terminal, and ``Wait before retrying.'' assumes a reader who can wait and then call again. The Model Context Protocol (MCP) puts a language-model agent in the reader's place. Many MCP servers are thin wrappers of such APIs. Of 116 official servers, 88.6\% are backed fully or partly by REST APIs, and 92\% implement their tools as bare API wrappers \cite{rest2mcp}.

An agent does not always have the developer's means. A coding agent in a terminal can run commands and wait between calls, but a chat application with MCP connectors, or an agent framework that gives its model only tool calls, can do nothing except call the tools of the connected servers. The server receives the same request from all of these callers and cannot tell which one it is answering. A step that is right for a developer is then out of reach for many of the agents that read it, and the one kind of step every caller can carry out is a call to one of the server's own tools.

The official guidance does not address this difference. The MCP specification describes errors in tool execution as feedback that lets the model correct itself and retry \cite{mcp}, and Anthropic's guide for tool authors recommends error responses that state specific, actionable improvements \cite{anthropictools}. Neither says who can act on the improvement, or reports how much a suggested step helps an agent recover. The newest models make the question more pressing. OpenAI and Anthropic report that GPT-6 Astra and Claude Fable 5 weigh written instructions more heavily than their predecessors, so that guidance written for earlier models can over-constrain them \cite{openaiastra,anthropicfable}.

We first measure how much MCP error text is written for a developer reader, in the source code of 150 widely used servers (Section~\ref{sec:survey}). We then measure what agents that act only through MCP tools do with that text. Each scenario replays a task from the Berkeley Function Calling Leaderboard (BFCL) \cite{bfcl} up to a tool call, makes the call fail, and continues from the same saved state with different error texts, and the BFCL state check decides whether the turn was completed. Last, we test two remedies, one for each side of the connection: the MCP developer can rewrite the step as a call to a server tool, and the agent developer, who cannot change third-party servers, can remove the step before the model reads it.

Half of the next steps in the surveyed error text depend on something the server cannot see about the caller, and on credential errors nearly every step asks for a terminal command, a configuration change or a web page. The agents did what these steps said. When the step asked for something outside their tools, most of them ended the turn, although their own tools could have made the repair, and the more capable the model, the more often it stopped, which is consistent with how OpenAI and Anthropic describe their newest models (Section~\ref{sec:discussion}). On expired credentials, a terminal command in the step left 45\% of tasks recovered; naming the login tool in its place raised this to 84\%, and deleting the step with a one-sentence prompt raised it to 82\%. On a rate limit, naming the call to repeat in place of ``Wait before retrying.'' raised recovery from 6\% to 88\%, at about one extra tool call. The paper contributes the survey, the controlled comparison across five models, the two remedies with their measured effect, and the public scenarios, data and code.

\section{Related Work}

\paragraph{Error messages for programmers and for agents} API design has long treated the developer as the reader: usability research studies how developers read documentation and error output and act on it \cite{myers2016}, and HTTP APIs now return errors in a standard machine-readable form that separates the problem type from its human-readable detail \cite{rfc9457}. Error messages have also been studied as text written for programmers, and compiler messages are often found unhelpful \cite{becker2019}. For coding agents, removing detail from type-error messages lowers the repair rate \cite{typeerror}, and tool feedback helps models revise their outputs \cite{critic,huang2024}. Tool-calling models often repeat a failed call, and training them to diagnose the failure improves recovery \cite{failurestronger}. We study a different property of the message, whether its reader can act on it.

\paragraph{Agents follow instructions in tool output} Agents carry out instructions an attacker places in tool output \cite{agentdojo,injecagent}, the most capable models most reliably follow instructions planted in MCP tool descriptions \cite{manuallies}, and larger models are more easily led by benign instruction-like sentences \cite{distractionif}. Agents also over-trust erroneous tool output \cite{toolrobust,toolmisuse}. The steps we study are written in good faith by the tool's author and are correct for a developer.

\paragraph{MCP servers} Studies of the MCP ecosystem cover faults \cite{taraghi2026}, defects in tool descriptions \cite{mcpsmells} and the wrapping of REST APIs \cite{rest2mcp}. We are aware of no study of the error text MCP servers return, or of what agents do with it.

\section{Error Text in Deployed MCP Servers}
\label{sec:survey}

We studied the 150 most-starred MCP servers on GitHub (426 to 186,635 stars, median 1,609) among those with public source that register their tools through an official MCP SDK and were updated in the past year. In each server we located the places where a failing tool returns text to the model, up to 30 per server, which gave 3,001 error messages. For each message we recorded whether it tells the caller what to do next and, if so, whether the step is correct only under conditions the server cannot observe, such as the caller's tools, login or permissions. OpenAI Codex agents assigned the labels from the source code, following a codebook released with the data \cite{artifacts}.

\begin{table}[t]
\centering
\small
\begin{tabular}{@{}lr@{}}
\toprule
Error messages & Count \\
\midrule
All messages in 150 servers & 3,001 \\
\quad with a next step & 949 \\
\quad\quad of which the step depends on the caller & 477 \\
Credential, permission and rate-limit messages & 209 \\
\quad with a next step & 128 \\
\quad\quad of which the step depends on the caller & 99 \\
\bottomrule
\end{tabular}
\caption{Next steps in the error messages of 150 widely used MCP servers}
\label{tab:survey}
\end{table}

Nearly every message states the cause of the failure, and 949 also tell the caller what to do next (Table~\ref{tab:survey}). Half of these steps depend on the caller. They are most common on failures the caller cannot see in its own call: of the 128 steps on credential, permission and rate-limit errors, 99 depend on the caller, and 93 of these ask for something a developer can do and an agent limited to MCP tools cannot, namely a configuration change (55), an action on a web page (24), a terminal command (13) or waiting (1). On credential errors, 62 of the 67 steps ask for a configuration change, a web page or a terminal command, as in ``Please run: \texttt{reddit-mcp-buddy -{}-auth}''. In 12 of these cases, from 5 servers, the server itself offers a tool that could make the repair, the situation the experiment reproduces. On rate limits, 20 of the 30 steps say to wait and retry, as in GitHub's ``Wait before retrying.'', and none says which call to repeat. These two are the steps the experiment tests.

\section{Experiment}

\subsection{Scenarios}

Scenarios come from the multi-turn tasks of BFCL V4 \cite{bfcl}, whose tools have executable implementations and whose turns have state-based success checks. We replay a task with its reference calls up to a chosen call, replace that call with a failing one, and save the environment state. There are 168 scenarios, 24 for each of seven failure types that a tool can diagnose when the call fails: a parameter in the wrong unit or format, a missing required field, a call to the wrong tool, expired credentials, a missing resource, a permission the session lacks, and a reached rate limit. Table~\ref{tab:settings} gives, for each type, how the call fails, the repair the environment accepts and the state by which recovery is judged.

\begin{table}[!t]
\centering
\footnotesize
\setlength{\tabcolsep}{4pt}
\begin{tabular}{@{}>{\raggedright\arraybackslash}p{0.15\linewidth}>{\raggedright\arraybackslash}p{0.29\linewidth}>{\raggedright\arraybackslash}p{0.25\linewidth}>{\raggedright\arraybackslash}p{0.22\linewidth}@{}}
\toprule
Failure type & How the call fails & Repair the environment accepts & Recovery is judged by \\
\midrule
Wrong unit or format & \texttt{fillFuelTank} receives the amount as text with a unit, such as ``40 gallons'' & the same call with a number & the fuel level \\
Missing required field & a required argument is left out, such as the directory name for \texttt{mkdir} & the call with the argument & the effect of the call \\
Wrong tool & the arguments of the intended call go to a listing or status tool of the same service, such as \texttt{ls} instead of \texttt{touch} & the intended tool with the same arguments & the effect of the intended call \\
Expired credentials & the service has no logged-in session, although the agent logged in earlier in the conversation & logging in with the earlier credentials, then the call & the logged-in user and the effect of the call \\
Missing resource & the call gives an identifier that no resource has (\texttt{-999999} or \texttt{\_\_missing\_resource\_\_}) for the resource the task names; the agent sees it in its own call & the call with the identifier of the named resource & the effect of the call on that resource \\
Missing permission & the session stays logged in, but the service rejects this one tool until a login to the same service succeeds & logging in with the earlier credentials, then the call & the effect of the call; the rejection itself is not compared \\
Rate limit reached & the call is rejected before it reaches the service, whose state is unchanged & the same call again, which succeeds at once; no wait is enforced & the effect of the call \\
\bottomrule
\end{tabular}
\caption{Setup of the seven failure types. Recovery means that at the end of the turn the state of every service equals its state after the turn's reference calls; the last column gives the part of that state that depends on the failed call.}
\label{tab:settings}
\end{table}

\subsection{Error text}

Each scenario has six texts in four conditions (Table~\ref{tab:conditions}). The generic notice is the same for all scenarios. The cause statement uses only information available to the tool when the call fails. The correct next step names an action that resolves the failure from the saved state, in two phrasings. The incorrect next step comes in two forms. An executable incorrect step names a tool the agent has, applied to the wrong target, so the action fails or moves the task away from its goal. An unavailable incorrect step asks for something the agent does not have, such as a terminal command, an environment variable or a tool outside its tool list. Next steps reuse the wording of real suggestions from the survey.

Two of the six texts reproduce the steps found in the survey, and two rewrite the same repair as a call to one of the server's tools. We call them the original step and the rewritten step. On expired credentials, the original step is the unavailable step, a terminal command, and the rewritten step is the second correct phrasing, which names the service's login tool. On a rate limit, the original step is the first correct phrasing, ``Wait before retrying.'', and the rewritten step is the second, which names the failed call, such as ``Wait a few seconds and call \texttt{place\_order} again''.

\begin{table}[t]
\centering
\small
\begin{tabular}{@{}l>{\raggedright\arraybackslash}p{0.64\linewidth}@{}}
\toprule
Condition & Text returned for scenario S016, expired credentials \\
\midrule
Generic & ``Operation failed.'' \\
Cause & ``\texttt{create\_ticket} rejected the call: no authenticated session exists.'' \\
Correct step, phrasing 1 & Cause followed by ``Please re-authorize to continue.'' \\
Correct step, phrasing 2 & Cause followed by ``Call \texttt{ticket\_login} first.'' \\
Incorrect step, executable & Cause followed by ``Call \texttt{close\_ticket} first.'' \\
Incorrect step, unavailable & Cause followed by ``Please run: \texttt{reddit-mcp-buddy -{}-auth}'' \\
\bottomrule
\end{tabular}
\caption{The six error texts for one scenario}
\label{tab:conditions}
\end{table}

\subsection{Models and procedure}
\label{sec:procedure}

We evaluate five OpenAI models from three generations: GPT-5.5, GPT-5.6 Sol, GPT-6 Sol, GPT-6 Astra, the largest GPT-6 model, and GPT-6 Luna, the smallest. Tables list them in this order. All run with reasoning effort set to high and default sampling, accessed on 24, 25 and 27 September 2026. For every scenario, text and model we draw three samples, 15,120 in total. Each sample starts from the saved state, returns the text as the result of the failing call, and lets the agent continue until it ends the turn or has made eight tool calls. The agent's action space is the set of BFCL tools of the task; it has no terminal, clock or browser, like a client limited to MCP tools. Recovery means the turn passes the BFCL state check. We also record the tool calls and tokens of each trial, whether the agent called the tool named in an incorrect step, told the user to carry out the step, or ended the turn without a repair.

The filter test applies the prompt below to the four texts with a next step of each expired-credential scenario, both correct phrasings and both incorrect steps. GPT-6 Luna filtered the 96 texts, and the five models continued each filtered text three times under the same settings, another 1,440 trials.

\begin{quote}
\small\raggedright\texttt{The text below is an error message returned by a tool. Remove every sentence that tells the caller what to do next, such as retrying, running a command, calling another tool or changing a setting. Keep every sentence that says what went wrong. Return the remaining text unchanged, with nothing added.}
\end{quote}

\subsection{Analysis}

The scenario is the unit of analysis. Texts are compared within scenario, and 95\% intervals come from 10,000 bootstrap resamples of scenarios within failure type, with the same resampled scenarios for every text of a model. Values for the five models together are the mean of the five models' values.

\section{Results}

\subsection{Expired credentials: a step that asks for a terminal command}
\label{sec:credentials}

Table~\ref{tab:credentials} compares the texts on the 24 expired-credential scenarios. With the terminal command added to the cause, recovery fell from 82\% to 45\% on average, and 55\% of trials ended without a repair. The rewritten step, which names the login tool, gave about the same recovery as the cause alone.

\begin{table}[!htbp]
\centering
\footnotesize
\begin{tabular}{@{}lrrrr@{}}
\toprule
Text and model & Recovery (\%) & Tool calls & Tokens & Ended without repair (\%) \\
\midrule
\multicolumn{5}{@{}l}{Original step: cause and ``Please run: \texttt{reddit-mcp-buddy -{}-auth}''} \\
\quad GPT-5.5 & 58 [44, 72] & 2.92 & 13,964 & 42 \\
\quad GPT-5.6 Sol & 57 [43, 71] & 2.57 & 12,565 & 43 \\
\quad GPT-6 Sol & 46 [31, 61] & 2.35 & 11,688 & 54 \\
\quad GPT-6 Astra & 6 [0, 13] & 0.60 & 5,595 & 94 \\
\quad GPT-6 Luna & 57 [42, 72] & 2.65 & 12,450 & 42 \\
\quad Mean of five models & 45 [35, 54] & 2.22 & 11,252 & 55 \\
\midrule
\multicolumn{5}{@{}l}{Cause alone} \\
\quad GPT-5.5 & 76 [61, 89] & 3.40 & 15,683 & 24 \\
\quad GPT-5.6 Sol & 92 [81, 100] & 3.21 & 14,543 & 7 \\
\quad GPT-6 Sol & 85 [71, 96] & 3.42 & 15,467 & 15 \\
\quad GPT-6 Astra & 75 [58, 92] & 2.78 & 13,182 & 25 \\
\quad GPT-6 Luna & 83 [71, 93] & 3.43 & 15,427 & 17 \\
\quad Mean of five models & 82 [70, 92] & 3.25 & 14,860 & 18 \\
\midrule
\multicolumn{5}{@{}l}{Rewritten step: cause and login tool, such as ``Call \texttt{ticket\_login} first.''} \\
\quad GPT-5.5 & 85 [71, 96] & 3.46 & 15,878 & 15 \\
\quad GPT-5.6 Sol & 89 [76, 99] & 2.64 & 12,809 & 11 \\
\quad GPT-6 Sol & 82 [69, 93] & 3.14 & 14,441 & 18 \\
\quad GPT-6 Astra & 75 [58, 92] & 1.99 & 10,420 & 25 \\
\quad GPT-6 Luna & 88 [75, 97] & 3.24 & 14,668 & 13 \\
\quad Mean of five models & 84 [72, 93] & 2.89 & 13,643 & 16 \\
\midrule
\multicolumn{5}{@{}l}{Original step, removed by the prompt run on GPT-6 Luna} \\
\quad GPT-5.5 & 86 [75, 96] & 3.88 & 16,954 & 11 \\
\quad GPT-5.6 Sol & 89 [75, 100] & 3.19 & 14,756 & 11 \\
\quad GPT-6 Sol & 76 [60, 90] & 3.47 & 15,723 & 24 \\
\quad GPT-6 Astra & 75 [58, 92] & 2.79 & 13,240 & 25 \\
\quad GPT-6 Luna & 85 [71, 96] & 3.21 & 14,534 & 15 \\
\quad Mean of five models & 82 [70, 93] & 3.31 & 15,041 & 17 \\
\bottomrule
\end{tabular}
\caption{Expired credentials, original and rewritten step: recovery with 95\% intervals, and tool calls, tokens and the share of trials that ended without a repair, per trial. Each text is averaged over the 24 scenarios and three runs per scenario.}
\label{tab:credentials}
\end{table}

The loss from the command grew with the model: 18 points for GPT-5.5, 35 for GPT-5.6 Sol, 39 for GPT-6 Sol and 69 for GPT-6 Astra, with 26 for the small GPT-6 Luna. The loss for GPT-6 Astra exceeded that for GPT-5.5 by 51 points (interval 29 to 72). Under the command, GPT-6 Astra made 0.60 tool calls per trial. In 48 of its 68 trials without a repair, its final message left the repair to the user: it asked the user to reconnect or sign in, said that the service needed authentication first, or pointed to the command, although the login tool was in its tool list.

Recovery took two calls, the login tool and the repeated call, with the credentials the agent had used earlier in the conversation. The rewritten step used fewer calls and tokens than the cause alone (Table~\ref{tab:credentials}).

\subsection{Rate limit: a step that says to wait}
\label{sec:ratelimit}

Table~\ref{tab:ratelimit} compares the two phrasings of the correct step on the 24 rate-limit scenarios. Under ``Wait before retrying.'', between 89 and 99\% of trials ended without a repair, depending on the model, and the mean trial made 0.37 tool calls. Naming the call to repeat raised recovery by 82 points (interval 75 to 88), to between 78 and 99\% per model. The mean trial then made 1.42 tool calls and used 5,409 tokens instead of 3,098.

\begin{table}[!htbp]
\centering
\footnotesize
\begin{tabular}{@{}lrrrr@{}}
\toprule
Text and model & Recovery (\%) & Tool calls & Tokens & Ended without repair (\%) \\
\midrule
\multicolumn{5}{@{}l}{Original step: ``Wait before retrying.''} \\
\quad GPT-5.5 & 4 [0, 11] & 0.25 & 2,899 & 96 \\
\quad GPT-5.6 Sol & 7 [0, 18] & 0.36 & 3,095 & 93 \\
\quad GPT-6 Sol & 8 [0, 19] & 0.50 & 3,326 & 92 \\
\quad GPT-6 Astra & 11 [1, 24] & 0.47 & 3,380 & 89 \\
\quad GPT-6 Luna & 1 [0, 4] & 0.26 & 2,791 & 99 \\
\quad Mean of five models & 6 [1, 13] & 0.37 & 3,098 & 94 \\
\midrule
\multicolumn{5}{@{}l}{Rewritten step: ``Wait a few seconds and call \emph{tool} again''} \\
\quad GPT-5.5 & 78 [65, 89] & 1.19 & 4,867 & 22 \\
\quad GPT-5.6 Sol & 99 [96, 100] & 1.39 & 5,357 & 1 \\
\quad GPT-6 Sol & 96 [88, 100] & 1.83 & 6,283 & 4 \\
\quad GPT-6 Astra & 89 [76, 100] & 1.44 & 5,643 & 11 \\
\quad GPT-6 Luna & 79 [69, 89] & 1.22 & 4,895 & 21 \\
\quad Mean of five models & 88 [83, 93] & 1.42 & 5,409 & 12 \\
\bottomrule
\end{tabular}
\caption{Rate limit, original and rewritten step: recovery with 95\% intervals, and tool calls, tokens and the share of trials that ended without a repair, per trial. Each text is averaged over the 24 scenarios and three runs per scenario.}
\label{tab:ratelimit}
\end{table}

\subsection{Removing the steps with a prompt}
\label{sec:filter}

GPT-6 Luna reduced all 96 texts to exactly the cause statement. On the text with the terminal command, filtering raised recovery by 37.5 points on average (interval 26.9 to 47.8), equal to recovery under the cause alone (Table~\ref{tab:credentials}). Where the removed step had been correct, recovery changed by $-$1.7 to $-$0.8 points, with every interval including zero.

On the 764 survey messages in which the step can be separated from the cause, the same prompt removed the step from 752 and left the rest of the message word for word in 548. Running it on all 949 messages with a step cost 0.09 US dollars.

\subsection{The other failure types}
\label{sec:other}

Table~\ref{tab:bytype} gives recovery for all seven failure types. When the failure lies in the agent's own call, a wrong unit or format, a missing field, a wrong tool or a missing resource, recovery under every text stayed within four points of that under the generic notice. On a missing permission and a reached rate limit, agents given only the cause ended the turn without a repair in at least 89\% of trials. There a step that named the repair was needed: naming the login tool raised recovery on a missing permission to 53\%, against 28\% for ``Please re-authorize to continue.''

\begin{table}[!htbp]
\centering
\small
\setlength{\tabcolsep}{4pt}
\begin{tabular}{@{}lrr>{\raggedleft\arraybackslash}p{2.8em}>{\raggedleft\arraybackslash}p{2.8em}rr@{}}
\toprule
 & & & \multicolumn{2}{c}{Correct step} & \multicolumn{2}{c}{Incorrect step} \\
\cmidrule(lr){4-5} \cmidrule(l){6-7}
Failure type & Generic & Cause & 1 & 2 & Executable & Unavailable \\
\midrule
Wrong unit or format & 83 & 79 & 82 & 82 & 81 & 81 \\
Missing required field & 86 & 88 & 88 & 88 & 87 & 86 \\
Wrong tool & 80 & 81 & 82 & 81 & 81 & 81 \\
Expired credentials & 61 & 82 & 84 & 84 & 81 & 45 \\
Missing resource & 98 & 98 & 97 & 97 & 99 & 98 \\
Missing permission & 0 & 0 & 28 & 53 & 0 & 0 \\
Rate limit reached & 59 & 5 & 6 & 88 & 66 & 8 \\
\bottomrule
\end{tabular}
\caption{Recovery (\%) by failure type, averaged over the five models. Columns 1 and 2 are the two phrasings of the correct step.}
\label{tab:bytype}
\end{table}

\section{Discussion}
\label{sec:discussion}

\paragraph{Why text written for developers fails these agents} A developer who reads ``Please run: \texttt{reddit-mcp-buddy -{}-auth}'' opens a terminal. An agent limited to MCP tools reads the same sentence and has no terminal. The agents in the experiment did what the step said: they carried out a step that named one of their tools, and when the step asked for something they could not do, most of them stopped, although their own tools could have made the repair and the same agents made it when the text gave only the cause. The text mattered only when the failure lay outside the agent's call. A malformed call shows the agent what to change whatever the message says, while a lapsed session, a missing permission or a rate limit is visible only through the text, and the agent then acts on what the text names. On a missing permission or a rate limit, stopping when given only the cause is the reading that HTTP gives a client for a 403 or a 429 response \cite{rfc9110,rfc6585}.

\paragraph{Why the newest model stops most often} Given only the cause, GPT-6 Astra logged in and recovered in most expired-credential trials; with one added sentence asking for a terminal command it almost never did, while the older and smaller models more often ignored the command and logged in (Table~\ref{tab:credentials}). We call this literal compliance: an agent that cannot carry out a suggested step gives up a repair its own tools could make. The model makers describe the change behind this in their newest models. OpenAI calls GPT-6 Astra its most aligned model and writes that it performs a task only when it knows the task is safe. Boundaries that developers wrote for earlier models can be taken too seriously, so that Astra stops where the developer would have wanted it to continue, and where GPT-5.6 Sol kept working through a request for long stretches, Astra can feel more tentative about when to stop and may come back for review while there is still work to do \cite{openaiastra}. Anthropic writes that skills developed for earlier models are often too prescriptive for Claude Fable 5 and can lower its output quality \cite{anthropicfable}, and its guide for Claude Fable 5.1 notes that the model sometimes describes what it would do next instead of doing it, so that the user has to reply before the work continues \cite{anthropicfable51}. Both descriptions are consistent with what we measured. A next step in an error message is an explicit written instruction. An earlier model tends to read past it and infer what the task needs; a model that weighs written instructions heavily and is quick to hand work back takes the step as the repair and, because the step lies outside its tools, leaves the repair to the user, which is what the final messages of GPT-6 Astra show (Section~\ref{sec:credentials}). The gap between GPT-5.6 Sol and GPT-6 Astra in Table~\ref{tab:credentials} is consistent with the difference the OpenAI guidance describes between the two models. Two studies from this year report the same direction for other text: larger models are more easily led by benign instruction-like sentences \cite{distractionif}, and the most capable models most reliably follow instructions planted in MCP tool descriptions \cite{manuallies}. The model makers address the problem for instructions that developers write for their own agents. The steps in error text are written by tool authors whom the agent's developer does not control, so the same server text can cost more recovery as agents move to newer models.

\paragraph{What an MCP developer can do} The server cannot tell whether its caller has a terminal, but every caller can call the server's tools. A step written as such a call works for a developer, a coding agent and an agent limited to MCP tools alike. Problem details for HTTP APIs give a program a machine-readable account of an error \cite{rfc9457}; a step that names a server tool gives the agent the part of that account it can act on, and in the experiment it gave the higher recovery on all three failures a server tool could repair (Tables~\ref{tab:credentials}, \ref{tab:ratelimit} and \ref{tab:bytype}). On a rate limit the call to repeat is always one of the server's own tools, so every server can write its step this way, and the essential part is the instruction to call again: the agents repeated the failed call even under a step that named the wrong tool. A command, a setting or a web page can still be offered to a person, but added to a credential error it lowered recovery below that of the cause alone for all five models.

\paragraph{What an agent developer can do} An agent developer usually connects servers written by others and cannot change their text. Where the server offers a tool that makes the repair, deleting the steps before the model reads them is cheap, raised recovery on expired credentials by 37.5 points, and cost nothing measurable where the removed step was correct (Section~\ref{sec:filter}). It suits credential errors, where the cause already points to the repair. On a missing permission or a rate limit the cause alone left most tasks unrecovered, so removing a correct step there would remove the repair.

\section{Threats to Validity}

Recovery is judged by the BFCL state check for the turn in which the failure occurs. The scenarios cover seven failure types across the BFCL multi-turn domains. The survey describes widely used open-source MCP servers on GitHub in September 2026. Model names are API aliases, and we report access dates.

\section{Conclusion}

MCP error text is often still written for the developer who used the API before MCP, and the agent that reads it now may only be able to call the server's tools. In 150 widely used MCP servers, most next steps on credential errors ask for a terminal command, a configuration change or a web page, and the steps on rate limits say to wait without naming the call to repeat. Agents limited to MCP tools did what these steps said and mostly stopped, and GPT-6 Astra, the largest model of the newest generation, stopped in almost every expired-credential trial. OpenAI and Anthropic describe their newest models as weighing written instructions more heavily, and OpenAI describes GPT-6 Astra as handing work back sooner, so the same server text can cost more recovery as agents move to newer models. Both sides of the connection can remove the problem. On expired credentials, recovery was 45\% with the terminal command, 84\% when the MCP developer named the login tool in its place, and 82\% when the agent developer deleted the step with a one-sentence prompt. On a rate limit, naming the call to repeat in place of the bare wait raised recovery from 6\% to 88\%.

\section*{CRediT authorship contribution statement}

\textbf{Xiaonan Xu:} Conceptualization, Methodology, Investigation, Writing -- original draft. \textbf{Wenjing Wu:} Software, Validation, Formal analysis, Writing -- review \& editing.

\section*{Declaration of competing interest}

We declare no competing financial interests or personal relationships that could have influenced this work.

\section*{Data availability}

Scenarios, error texts, survey data, model outputs and code are available at \url{https://github.com/WenJing95/tool-error-text}.

\sloppy

\end{document}